\documentclass[apj,tighten,iop,twocolumn]{emulateapj}

\shorttitle{Faint Quasar at $z\sim6$}
\shortauthors{Kim et al.}

\begin{document}

\title{Discovery of A Faint Quasar at $z \sim 6$ and Implications for Cosmic Reionization}

\author{Yongjung Kim\altaffilmark{1,2}, 
Myungshin Im\altaffilmark{1,2}, 
Yiseul Jeon\altaffilmark{1,2},
Minjin Kim\altaffilmark{3,4},
Changsu Choi\altaffilmark{1,2}, 
Jueun Hong\altaffilmark{1,2}, 
Minhee Hyun\altaffilmark{1,2}, 
Hyunsung David Jun\altaffilmark{1,5}, 
Marios Karouzos\altaffilmark{2}, 
Dohyeong Kim\altaffilmark{1,2}, 
Duho Kim\altaffilmark{1,6}, 
Jae-Woo Kim\altaffilmark{1,2}, 
Ji Hoon Kim\altaffilmark{7}, 
Seong-Kook Lee\altaffilmark{1,2}, 
Soojong Pak\altaffilmark{8},
Won-Kee Park\altaffilmark{3}, 
Yoon Chan Taak\altaffilmark{1,2}, 
and Yongmin Yoon\altaffilmark{1,2}
}
\email{yjkim@astro.snu.ac.kr \& mim@astro.snu.ac.kr}

\altaffiltext{1}{Center for the Exploration of the Origin of the Universe (CEOU), Building 45, Seoul National University, 1 Gwanak-ro, Gwanak-gu, Seoul 151-742, Korea}
\altaffiltext{2}{Astronomy Program, FPRD, Department of Physics \& Astronomy, Seoul National University, 1 Gwanak-ro, Gwanak-gu, Seoul 151-742, Korea}
\altaffiltext{3}{Korea Astronomy and Space Science Institute, Daejeon 305-348, Korea}
\altaffiltext{4}{University of Science and Technology, Daejeon 305-350, Korea}
\altaffiltext{5}{Jet Propulsion Laboratory, California Institute of Technology, 4800 Oak Grove Drive, Pasadena, CA 91109, USA}
\altaffiltext{6}{Arizona State University, School of Earth and Space Exploration, P.O. Box 871404, Tempe, AZ 85287-1404, U.S.A.}
\altaffiltext{7}{Subaru Telescope, National Astronomical Observatory of Japan, 650 North A'ohoku Place, Hilo, HI 96720, U.S.A.}
\altaffiltext{8}{School of Space Research and Institute of Natural Sciences, Kyung Hee University, 1732 Deogyeong-daero, Giheung-gu, Yongin-si, Gyeonggi-do 446-701, Korea}

\submitted{Received 2015 September 4; accepted 2015 September 27; published 2015 November 5} 

\begin{abstract}
Recent studies suggest that faint active galactic nuclei may be responsible for the reionization of the universe.
Confirmation of this scenario requires spectroscopic identification of faint quasars ($M_{1450}>-24$ mag) at $z \gtrsim6$, but only a very small number of such quasars have been spectroscopically identified so far.
Here, we report the discovery of a faint quasar IMS J220417.92+011144.8 at $z\sim6$ in a 12.5 deg$^{2}$ region of the SA22 field of the Infrared Medium-deep Survey (IMS).
The spectrum of the quasar shows a sharp break at $\sim8443~\rm{\AA}$, with emission lines redshifted to $z=5.944 \pm 0.002$ and rest-frame ultraviolet continuum magnitude $M_{1450}=-23.59\pm0.10$ AB mag.
The discovery of IMS J220417.92+011144.8 is consistent with the expected number of quasars at $z\sim6$ estimated from quasar luminosity functions 
based on previous observations of spectroscopically identified low-luminosity quasars .
This suggests that the number of $M_{1450}\sim-23$ mag quasars at $z\sim6$ may not be high enough to fully account for the reionization of the universe. 
In addition, our study demonstrates that faint quasars in the early universe can be identified effectively with a moderately wide and deep near-infrared survey such as the IMS. 
\end{abstract}


\section{Introduction}\label{sec:intro}

Several dozens of quasars are now identified at $z \gtrsim 6$ \citep{Fan06,Goto06,Willott10b,Mortlock11,Venemans13,Banados14,Kashikawa15,Venemans15,Wu15}.
They are found to be powered by supermassive black holes (SMBHs) with masses of $10^{8}-10^{10}~M_{\odot}$ \citep{Jiang07,Willott10a,Mortlock11,Jun15}, shining at the Eddington limit, meaning that they are accreting mass at their maximal rates \citep{Willott10a,Jun15}, and some of them show paucity of hot dust emission \citep{Jiang10,Jun13} in contrast to quasars at low redshift \citep{KimD15}.
These results suggest that high-redshift quasars already harbor SMBHs at their centers just $\sim1$ Gyr after the Big Bang, and these SMBHs are growing more vigorously than their counterparts at low redshift. 
Furthermore, strong Gunn-Peterson troughs \citep{Gunn65} in their spectra suggest that a significant fraction of the intergalactic medium (IGM) is reionized at $z\sim6$ \citep{Fan06}.
However, two interesting questions still remain due to the lack of known faint quasars at $z \gtrsim 6$ ($M_{1450} > -24$ mag).

First, while the bright end of the $z\sim6$ quasar luminosity function, which is not sufficient to reionize the IGM, is well constrained, the faint end of the function is still debated.
Recently, \cite{Giallongo15} found 22 faint active galactic nuclei (AGNs) candidates with X-ray detections at $z>4$, indicating that there are more faint AGNs than previously expected, and the faint AGNs could be main contributors to the reionization of the universe \citep{Glikman11,Madau15}.
Unfortunately, the number of 
spectroscopically confirmed $z\sim6$ faint quasars is very small (only 
a few; \citealt{Willott09,Kashikawa15}). 
Consequently, the faint end of the quasar luminosity function is still very uncertain, and the potential role of this population in the reionization of the early universe is yet unclear.

Second, most of the $z\sim6$ quasars discovered so far are luminous, $M_{1450} < -24.5$ mag ($z'<22$ mag), implying that the currently discovered sample of high-redshift quasars is biased to objects with high accretion rates.
Such a bias hinders our understanding of the range of accretion rates of quasars in the early universe. 

Recently, we have been performing the Infrared Medium-deep Survey (IMS), a moderately wide (120 deg$^{2}$), and moderately deep ($J \sim 22.5 - 23$ mag) near-infrared (NIR) imaging survey with the Wide Field Camera (WFCAM; \citealt{Casali07}) on the United Kingdom InfraRed Telescope (UKIRT).
One of the main scientific aims of the IMS is to discover faint quasars at $z\sim6$ and beyond.
We combined the $Y$- and $J$-band imaging data from the IMS with optical data from the Canada-France-Hawaii Telescope Legacy Survey (CFHTLS; \citealt{Hudelot12})
and other Canada-France-Hawaii Telescope (CFHT) PI programs,
and we employed multiple color selection criteria to identify $z\sim6$ quasar candidates.
Here, we present the discovery of the first IMS faint quasar at $z\sim6$, and we discuss how this can constrain the sources responsible for the reionization of the universe.

In this Letter, we assumed the cosmological parameters of $\Omega_{m} =0.3$, $\Omega_{\Lambda}=0.7$ (e.g., \citealt{Im97}), and $H_{0}=70$ km s$^{-1}$ Mpc$^{-1}$.
All magnitudes are given in the AB system.

\section{IMS and CFHTLS Data}\label{sec:data}

The IMS includes seven extragalactic fields (XMM-LSS, CFHTLS-W2, Lockman Hole, EGS, ELAIS-N1, ELAIS-N2, and SA22; Im, M. et al. 2015, in preparation).
In this Letter ,we focus on one of the IMS fields, SA22, which covers $\sim 12.5~$deg$^{2}$ on the CFHTLS Wide 4 field, centered at $\alpha=22^{\rm{h}} 13^{\rm{m}} 18^{\rm{s}}$, $\delta=+01^{\circ} 19'00''$ (J2000).
The SA22 field images have 5$\sigma$ detection limits for a point source with 2$\times$FWHM aperture magnitudes of 23.5 and 23.3 mag in the $Y$- and $J$-bands, respectively (the median seeing values are $0\farcs93$ and $0\farcs84$, respectively; for details, see \citealt{Karouzos14}).
The $J$-band depth corresponds to $M_{1450}\simeq-23$ mag for a quasar at $z=6$, which is 2 mag fainter than most luminous $z\sim6$ quasars discovered to date.
The magnitude limits in the optical bands from CFHTLS are $u^{*} \sim25.2$, $g'\sim25.5$, $r'\sim25.0$, $i'/y'\sim24.8$, and $z'\sim23.9$ mag at the 80\% completeness limit for point sources \citep{Hudelot12}. 

Source detection was performed in $z'$-band images of CFHTLS using SExtractor \citep{Bertin96}.
DETECT$\_$THRESH and DETECT$\_$MINAREA parameters were set as 1.1 and 7, respectively. 
Using the identified $z'$-band sources, we measured fluxes at bluer bands using the dual mode of the SExtractor software.
Aperture magnitudes with $2 \times$FWHM$_{z'}$ diameters (FWHM$_{z'}\sim0\farcs7$) were used to measure fluxes\footnote{The signal-to-noise ratio (S/N; FLUX/FLUXERR from SExtractor) of point source peaks for apertures with diameters of $\sim2\times$FWHM$_{z'}$. }$^{,}$\footnote{
The errors are based on pixel-to-pixel noise of our stacked images made of sub-pixel dithered frames. We find that this can underestimate the actual photometric error by $\sim$ 20 -- 30 \% (e.g., \citealt{Jeon10}). 
}.
The aperture fluxes were converted to total fluxes by applying aperture corrections derived using bright stars in each filter image. 

For the IMS NIR images, we detected the sources using the association mode of the SExtractor software.
We matched sources in the IMS $Y$- and $J$-band catalogs with those in the $z'$-band catalog using a matching radius of $0\farcs5$.
The fluxes in the $Y$- and $J$-bands were obtained in a similar way to the CFHTLS data.

\begin{figure}
\epsscale{1.20}
\plotone{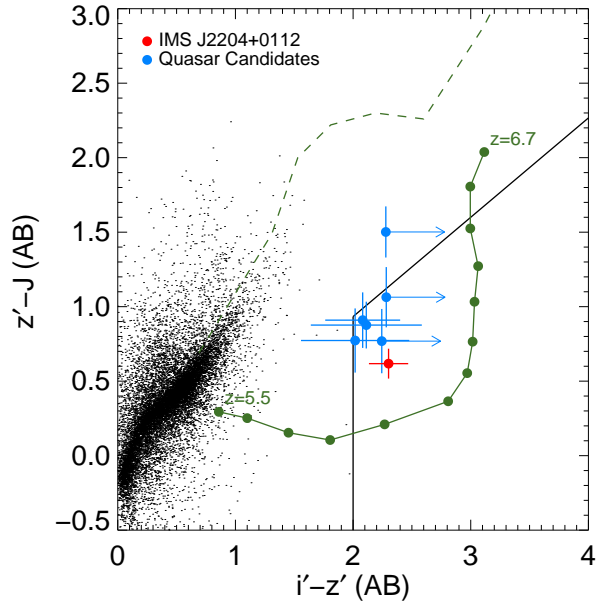}
\caption{Color-color diagram to identify quasar candidates at $z\sim6$. The red point represents the newly discovered $z\sim6$ quasar, IMS J2204+0112, while the blue points are other quasar candidates at $z\sim6$. The green points and solid line represent a quasar evolution track at $5.5<z<6.7$ in bins of 0.1, while the green dashed line is the mean color distribution of M/L/T dwarfs, which need to be separated from quasar candidates \citep{Willott05}. The black dots are sources classified as isolated point sources (FLAG$=$0, stellarity $>$ 0.8) in a one-twelfth area of the IMS SA22 field.  \label{fig:ccd}}
\end{figure}

\section{Quasar Candidate Selection}\label{sec:sel}

The spectral energy distributions (SEDs) of quasars at $z\sim 6$ show a sudden break blueward of Ly$\alpha$ (rest-frame 1216 $\rm{\AA}$) due to the Gunn-Peterson effect \citep{Gunn65}.
The break is located at $\lambda \sim 8500~\rm{\AA}$ for a quasar at $z \sim 6$, which creates a very red $i'-z'$ color, but blue colors at longer wavelengths.
Figure \ref{fig:ccd} shows the color-color diagram for point sources in the IMS SA22 field.
We used color cuts of $(i'-z') >2$ and $(i'-z')-1.5(z'-J)>0.6$ for sources at $z'<z'_{7\sigma}$ and required no detection (2$\sigma$ level) in $u'$-, $g'$-, and $r'$-band images. Here, $z'_{7\sigma}$ is the 7$\sigma$ detection limit in the $z'$-band. 
The first color cut samples objects with a strong break at $\lambda \sim 8500 ~ \rm{\AA}$, and the second color cut removes red stars such as brown dwarfs.
These color selection criteria are identical to those in \cite{Willott09}.

Spurious detections such as diffraction spikes could produce a number of false $i'$-dropout objects with very red $i'-z'$ colors that satisfy the color selection criteria.
To remove spurious sources, we reject objects that (1) have FLAG$\neq0$ in the SExtractor-produced catalogs to avoid blended objects and (2) are located at the edge of each image.
The remaining objects were visually inspected 
in all filter images 
to finalize the quasar candidate selection (blue points in Figure \ref{fig:ccd}).
Most of the sources removed by visual inspection are diffraction spikes, crosstalks, or other image artifacts.
Through this process, we identified seven quasar candidates at $z\sim6$ in the IMS SA22 field.

\begin{deluxetable*}{ccccccc}
\tabletypesize{\scriptsize}
\tablecaption{Properties of IMS J220417.92+011144.8\label{tbl:info}}
\tablewidth{500pt}
\tablehead{
\colhead{R.A.  Decl. (J2000.0)} & 
\colhead{$i'$} & \colhead{$z'$} & \colhead{$Y$} & \colhead{$J$} & 
\colhead{Redshift} & \colhead{$M_{1450}$}
}
\startdata
22:04:17.92 +01:11:44.8 & 
$25.26\pm0.15$ & $22.95\pm0.07$ & $23.10\pm0.09$ & $22.34 \pm 0.08$ & 
$5.944\pm0.002$ & $-23.59\pm0.10$
\enddata
\end{deluxetable*}

\begin{figure*}
\epsscale{1.0}
\plotone{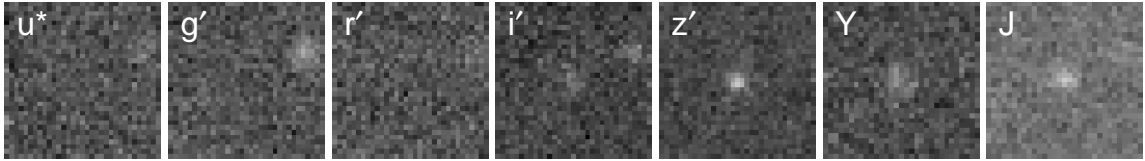}
\caption{Postage stamp images of IMS J2204+0112 presented in $6''\times6''$ boxes. From left to right, $u^{*}$-, $g'$-, $r'$-, $i'$-, $z'$-, $Y$-, and $J$-band images are shown. \label{fig:stamp}}
\end{figure*}

\begin{figure*}
\epsscale{0.9}
\plotone{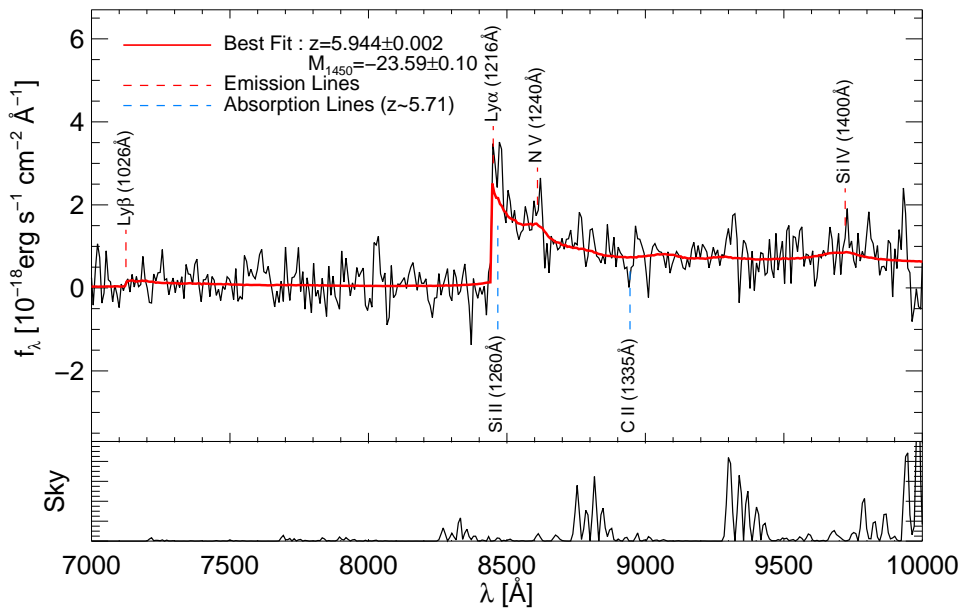}
\caption{Optical spectrum of IMS J2204+0112 (black line). The red line represents an SDSS quasar composite spectrum \citep{Vanden01} that is redshifted to $z=5.944$ and adjusted for IGM attenuation \citep{Madau96}. Quasar emission lines (red dashed) and possible absorption lines from a foreground source at $z\simeq5.71$ (blue dashed) are also indicated. The bottom figure shows skylines.\label{fig:spec}}
\end{figure*}

\section{Spectroscopic Observation}\label{sec:spec}

Among the seven quasar candidates at $z\sim6$, IMS J220417.92+011144.8 (hereafter IMS J2204+0112) was given the highest priority for follow-up spectroscopy due to small magnitude errors ($\Delta z' \sim 0.07$, $\Delta J \sim 0.08$; see Table \ref{tbl:info}), the location of the object in the selection box, and its point-like shape in $z'$- and $J$-band images (see Figure \ref{fig:stamp}). 
We observed this target with the Gemini Multi-Object Spectrograph (GMOS; \citealt{Hook04}) on the 8m Gemini South Telescope in Chile, on 2015 July 21 and 23 (Program ID: GS-2015A-Q-201).

The observation was carried out using the Nod \& Shuffle (N\&S) longslit mode with the R150\_G5326 grating to facilitate skyline subtraction.
We set the central wavelength to $9000~\rm{\AA}$ to avoid hot columns and spurious charges in one of the CCD chips that had a technical problem.
Since we wanted to cover the gap between each CCD chip, the observation was made with two configurations of the grating with central wavelengths of $8900$ and $9000~\rm{\AA}$.
We adopted the N\&S slit with $1\farcs0$ width and $4\times4$ pixel binning for maximal S/N, which gives a spectral resolution of $7.72~\rm{\AA}~\rm{pixel}^{-1}$ ($\sim290$ km s$^{-1}$).
In addition, the RG610\_G0331 filter was used to avoid the order-overlap.
Each N\&S sequence contained eight cycles of $60$ s exposure and together with overheads, lasted 968 s.
Although we observed 12 sequences for IMS J2204+0112, which gives a total exposure time of $\sim 3$ hr, we opted to use only five frames that were taken during good weather conditions (seeing $\lesssim1\farcs0$, gray night).

We followed the standard data reduction procedure with the IRAF/Gemini package: (1) bias subtraction and flat-fielding, (2) sky subtraction with shuffled spectra, (3) wavelength calibration with the CuAr arc lines, and (4) flux calibration with a spectrophotometric star (LTT7987).
After the flux calibration, we adjusted the overall flux scale using the photometric magnitude in the $z'$-band.

\section{Discovery of A Faint Quasar at $z\sim6$}\label{sec:disc}

Figure \ref{fig:spec} shows the spectrum of IMS J2204+0112. It shows a clear break at $\sim8443~\rm{\AA}$ that can be identified as the redshifted Ly$\alpha$ line. 
We fit the spectrum with a composite spectrum of SDSS quasars \citep{Vanden01}, including the IGM attenuation \citep{Madau96}.
The fit with the composite spectrum matches the observed spectrum of IMS J2204+0112 well and gives a redshift of $z=5.944\pm0.002$.
We estimate the absolute magnitude at the rest-frame wavelength of $1450~\rm{\AA}$ from the quasar spectrum to be $M_{1450}=-23.59\pm0.10$ mag.
IMS J2204+0112 shows strong Ly$\alpha$ and \ion{N}{5} (rest-frame 1240 $\rm{\AA}$) emission lines at $\sim 8600~\rm{\AA}$, while a weak \ion{Si}{4} (1400 $\rm{\AA}$) line can be seen at $\sim 9700~\rm{\AA}$.
Additionally, we identify possible absorption lines of \ion{Si}{2} (1260 $\rm{\AA}$) and \ion{C}{2} (1335 $\rm{\AA}$) at $\sim8460$ and $\sim8940~\rm{\AA}$, respectively, corresponding to an absorber at $z\sim5.71$.
The absolute magnitude of $M_{1450}=-23.59$ mag ranks IMS J2204+0112 as the third or the fourth faintest quasar at $z\sim6$ discovered to date \citep{Willott09,Kashikawa15}, depending on whether we treat the faintest quasar in \cite{Kashikawa15} as a quasar or a Lyman break galaxy.

If we assume that the quasar is accreting at the Eddington limit ($\lambda_{\rm{Edd}}=1$), then the black hole mass of the quasar is $M_{\rm{BH}}\sim10^{8}~M_{\odot}$ \citep{Vestergaard09,Jun15}.
We take this values as a lower limit of $M_{\rm{BH}}$ since this object could have $\lambda_{\rm{Edd}}$ lower than 1.

\begin{figure}
\epsscale{1.2}
\plotone{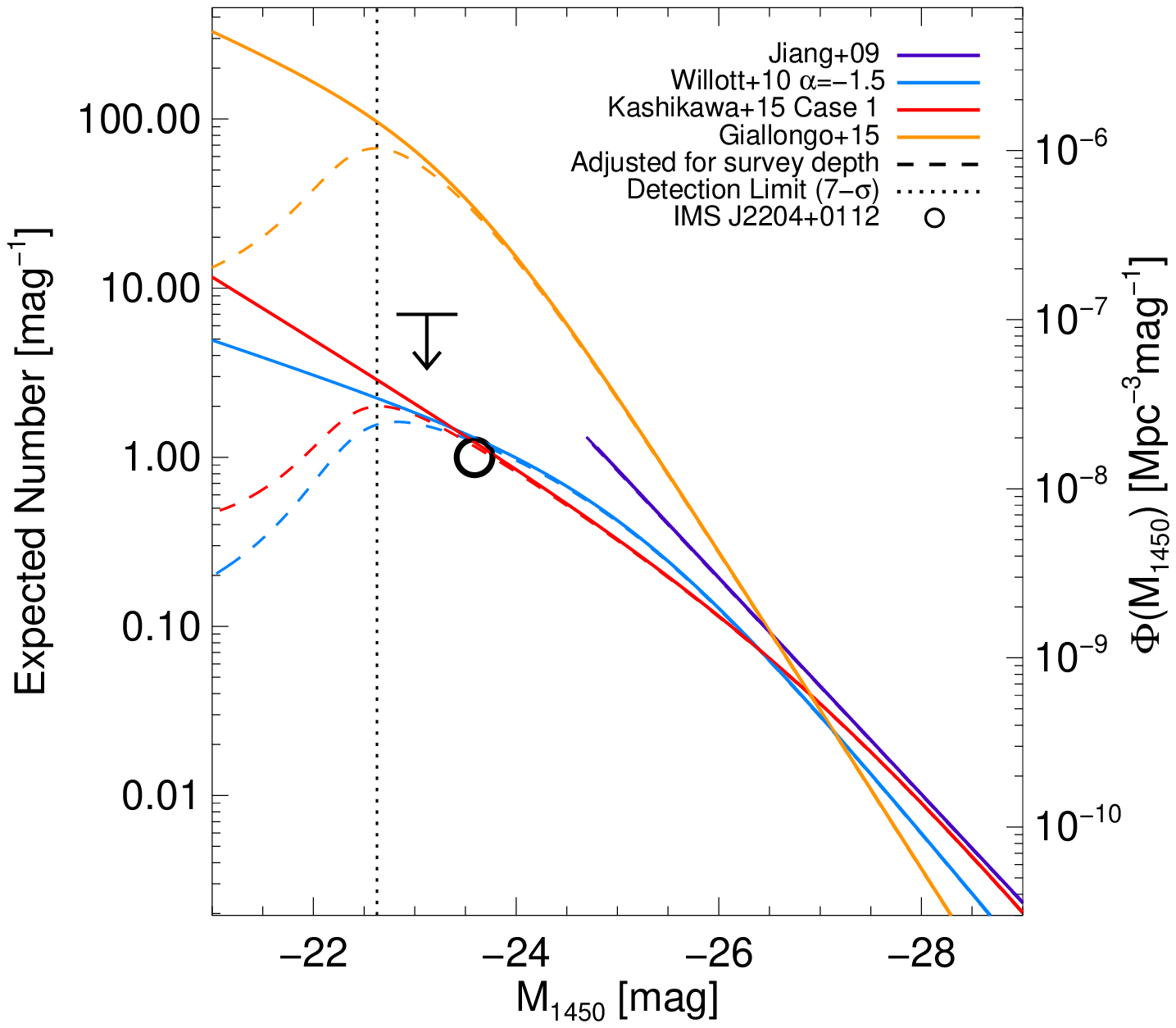}
\caption{Expected numbers of $z\sim6$ quasars in a 12.5 deg$^{2}$ area, inferred from $z\sim6$ quasar luminosity functions reported in the literature. The purple line is the single power-law luminosity function from \cite{Jiang09}, while the blue, red, and orange lines are double power-law luminosity functions from \cite{Willott10b}, \cite{Kashikawa15}, and \cite{Giallongo15}, respectively. The dashed lines represent the expected numbers after the completeness correction, and the vertical dotted line shows the CFHTLS 7$\sigma$ detection limit in the $z'$-band. IMS J2204+0112 is shown with the open circle, while the upper limit of the number that includes the additional six candidates for quasars at $z\sim6$ is represented with a downward arrow. \label{fig:qlf}}
\end{figure}

\section{Discussion}\label{sec:diss}

We can ask how the discovery of a $z\sim6$ quasar with $M_{1450}=-23.5$ mag constrains the faint end of the quasar luminosity function, a question that is directly related to the reionization of the universe (Section \ref{sec:intro}).
However, since we cannot robustly constrain the luminosity function using only one object, we instead estimate how many $z\sim6$ faint quasars are expected in a 12.5 deg$^{2}$ area and see if the number matches with our discovery.

To measure the expected number of $z\sim6$ quasars, we estimate the completeness in the $z'$-band, as we detected the quasar candidates in $z'$-band.
Using a simple minimum chi-square method, we fit the completeness data for $z'$-band of CFHTLS \citep{Hudelot12} with an analytic completeness function of the form $f$ \citep{Fleming95}:
\begin{equation}
f=\frac{1}{2}\left( 1 - \frac{\alpha (z'-z'_{*})}{\sqrt{(1+\alpha^{2} (z'-z'_{*})^{2})}}\right)
\end{equation}
where $z'_{*}$ is the turnover magnitude at which $f$ reaches 0.5 and $\alpha$ is the slope at $z'_{*}$.
We obtained $z'_{*}=24.26$ mag and $\alpha=1.7$.

Figure \ref{fig:qlf} shows the expected number (or quasar luminosity function) of $z\sim6$ quasars over an area of 12.5 deg$^{2}$.
Note that \cite{Giallongo15} presented their quasar luminosity function at $z=5.75$.
We scaled down this luminosity function by a factor of $10^{-0.47(5.75-6.0)}$ to take into account the number density evolution \citep{Willott10b} between $z=5.75$ and 6.0.
Then we evaluated the number density by integrating the luminosity function over a redshift interval of $5.7<z<6.3$ that roughly corresponds to our selection window in redshift space.
The solid lines represent the expected numbers from various quasar luminosity functions \citep{Jiang09,Willott10b,Giallongo15,Kashikawa15}, while the dashed lines show the cases adjusted for the survey depth with the completeness function.
The purple line represents a single power-law luminosity function up to $M_{1450}\sim-25$ mag with a slope of $-2.6$ \citep{Jiang09}, while the blue and red lines represent double power-law luminosity functions from \cite{Willott10b} and \cite{Kashikawa15}, respectively.
The quasar luminosity function of \cite{Giallongo15} is also shown as an orange line.
The vertical dotted line represents the 7$\sigma$ detection limit in the $z'$-band.
IMS J2204+0112 is plotted with an open circle. The error bar of the quasar is not shown because of the large uncertainty due to the small number statistics.
The downward arrow represents the upper limit of the expected number that includes IMS J2204+0112 and the other six candidates for quasar at $z\sim6$.
The average absolute magnitude of the seven sources is $M_{1450}=-23.12\pm0.30$ mag.

For both luminosity functions of \cite{Willott10b} and \cite{Kashikawa15}, the expected number of quasars is $1.4$ at $M_{1450}=-23.5$ mag, which is consistent with the number of $z\sim6$ faint quasars we discovered so far.
If we include the other six candidates in the comparison (downward arrow in Figure \ref{fig:qlf}), the number becomes $\sim5$ times higher than the expected numbers from the luminosity functions \citep{Willott10b,Kashikawa15}.
On the other hand, with the quasar luminosity function of \cite{Giallongo15}, the expected number of quasars at $z\sim6$ is $\sim40$ at $M_{1450}\sim-23.5$ mag, which is significantly higher than our result even if we consider the spectroscopically unconfirmed candidates as real quasars.
\cite{Giallongo15} noted that AGN populations near $M_{1450}\sim-23.5$ contribute the most to the AGN hydrogen-ionizing emissivity.
With their luminosity function, we find that AGN populations near the magnitude $M_{1450}\sim-23.5$ ($-25<M_{1450}<-22$) would emit about 60 \% of the required ultraviolet photons for the cosmic reionization.
However, the emissivity estimated with luminosity functions of \cite{Willott10b} and \cite{Kashikawa15} at that magnitude interval is only $\sim3$ \% of the required emissivity.
If faint quasars at $z\sim6$ are as abundant as the upper limit of our result, the fraction of the emissivity at $M_{1450}\sim-23.5$ ($-25<M_{1450}<-22$) reaches up to $\sim15~\%$ of the required emissivity of the IGM reionization.
The existence of IMS J2204+0112 is more in line with the quasar luminosity functions at $z\sim6$ of \cite{Willott10b} and \cite{Kashikawa15} rather than from \cite{Giallongo15}.
 Note that \cite{Giallongo15} interpolated between bright quasars ($M_{1450} < -25$) and 
faint quasar candidates ($-21 < M_{1450} < -19$) with photometric redshifts 
to derive the quasar luminosity function at $M_{1450} \sim -23$, while the 
luminosity functions of \cite{Willott10b} and \cite{Kashikawa15} are based
on a small number of spectroscopically confirmed quasars.
However, with the six quasar candidates, there is a possibility that the contribution of the quasars at $z\sim6$ to the cosmic reionization can be higher than previous predictions, but they might still not be the main contributors.

Our discovery of IMS J2204+0112 verifies that a moderately deep and wide NIR survey such as the IMS can efficiently discover low-luminosity quasars at $z\sim6$.
We expect that the analysis of the full IMS data set will net 10 times more faint quasars at $z\sim6$ than this study, supplying a sufficiently large sample of low-luminosity quasars to constrain the faint end of the quasar luminosity function and help understand the contribution of high-redshift quasars to the reionization of IGM in the early universe.
Such a sample, combined with deep NIR spectroscopy, will also allow us to determine black hole masses and Eddington ratios of the faint quasar population. Ultimately, we can gain a more complete view of the quasar accretion activity at $z\sim6$.

\acknowledgments

This work was supported by the National Research Foundation of Korea (NRF) grant No. 2008-0060544, funded by the Korea government (MSIP).
This work was supported by K-GMT Science Program (PID: gemini\_KR-2015A-023) of Korea Astronomy and Space Science Institute (KASI).
Based on observations obtained with MegaPrime/MegaCam, a joint project of CFHT and CEA/IRFU, at the Canada-France-Hawaii Telescope (CFHT), which is operated by the National Research Council (NRC) of Canada, the Institut National des Science de l'Univers of the Centre National de la Recherche Scientifique (CNRS) of France, and the University of Hawaii. 
This work is based in part on data products produced at Terapix available at the Canadian Astronomy Data Centre as part of the Canada-France-Hawaii Telescope Legacy Survey, a collaborative project of NRC and CNRS.
The United Kingdom Infrared Telescope (UKIRT) is supported by NASA and operated under an agreement among the University of Hawaii, the University of Arizona, and Lockheed Martin Advanced Technology Center; operations are enabled through the cooperation of the Joint Astronomy Centre of the Science and Technology Facilities Council of the UK.
Based on observations obtained at the Gemini Observatory acquired through the Gemini Science Archive and processed using the Gemini IRAF package, which is operated by the Association of Universities for Research in Astronomy, Inc., under a cooperative agreement with the NSF on behalf of the Gemini partnership: the National Science Foundation 
(United States), the National Research Council (Canada), CONICYT (Chile), the Australian Research Council (Australia), Minist\'{e}rio da Ci\^{e}ncia, Tecnologia e Inova\c{c}\~{a}o (Brazil) and Ministerio de Ciencia, Tecnolog\'{i}a e Innovaci\'{o}n Productiva (Argentina).
H.D.J. is supported by an appointment to the NASA Postdoctoral Program at the Jet Propulsion Laboratory, administered by Oak Ridge Associated Universities through a contract with NASA.
D.K. acknowledges the fellowship support from the grant NRF-2015-Fostering Core Leaders of Future Program No. 2015-000714 funded by the Korean government.
M.H. acknowledges support from the Global Ph.D. Fellowship Program through the National Research Foundation of Korea (NRF) funded by the Ministry of Education (NRF-2013H1A2A1033110).

{\it Facilities:} \facility{UKIRT (WFCAM)}, \facility{CFHT (MegaCam)}, \facility{Gemini:South (GMOS-S)}

\end{document}